\documentstyle[12pt]{article}
\newcommand{\beq}{\begin{equation}}
\newcommand{\eeq}{\end{equation}}
\newcommand{\bea}{\begin{eqnarray}}
\newcommand{\eea}{\end{eqnarray}}

\newcommand{\goto}{\rightarrow}

\newcommand{\D}{\Delta}

\begin{document}
\topmargin 0pt
\oddsidemargin 5mm
\renewcommand{\thefootnote}{\fnsymbol{footnote}}
\newpage
\setcounter{page}{0}
\begin{titlepage}
\begin{flushright} 
{\tt hep-th/9911163
}
 \end{flushright}
\bigskip
\bigskip
 
\begin{center}
{\Large Logarithmic Operators in $\mbox{AdS}_3/\mbox{CFT}_2$}
\bigskip 
\bigskip

{Alex Lewis}
\bigskip
{Department of Mathematical Physics, National University of Ireland,
Maynooth, Republic of Ireland. e-mail: alex@thphys.may.ie}

\end{center}
\begin{center}
\footnotesize
        
	\end{center}               

\normalsize 
\bigskip 
\bigskip
\begin{center}
			{\bf Abstract}
\end{center}
{We discuss the relation between singletons in
$\mbox{AdS}_3$  and logarithmic
operators in the CFT on the boundary. In $2$ dimensions
there can be more logarithmic operators apart from 
those which correspond to
singletons in AdS, because logarithmic operators can occur when the
dimensions of primary fields differ by an integer instead of being
equal. These operators may be needed to account for the greybody
factor for gauge bosons in the bulk.}
\end{titlepage}
\newpage


One particularly interesting example of the AdS/CFT correspondence
\cite{maldacena,gkp,witten} is the $\mbox{AdS}_3/\mbox{CFT}_2$ correspondence, which relates
supergravity on $\mbox{AdS}_3 \times S^3$ to a 2-dimensional CFT. One
advantage of this is that 2-dimensional conformal field theories are
very well understood, and that makes $\mbox{AdS}_3$ especially suitable for
studying the relation between singletons on AdS and logarithmic
conformal field theories (LCFT), since almost all previous work on
LCFT has concentrated on the 2-dimensional case. 

According to \cite{gkp,witten}, at the boundary of $\mbox{AdS}_{D+1}$ we have a
coupling between bulk fields $\Phi_i(\vec{x},z)$ and boundary fields
$O_i(\vec{x})$, $\int d^Dx \Phi_iO_i$, where the boundary fields are
subject to the boundary condition
$\Phi_i(\vec{x},z)=\lambda_i(\vec{x},R)$, with $z=R$ the boundary of
$\mbox{AdS}_{d+1}$. The relation between  
correlation functions in $CFT_D$ and the bulk supergravity action is
\beq
\langle e^{\sum_i\int d^Dx \lambda_i O_i} \rangle =
e^{-S[\{\Phi_i\}]}
\label{adscft}\eeq
This relation was used in \cite{gka,kogan} to show that, if there are
singletons in $\mbox{AdS}_{d+1}$, the theory on the boundary is in fact an
LCFT. A theory of free singletons is formulated in terms of a
dipole-ghost pair of fields $A$ and $B$ which satisfy \cite{ff}
\beq
(\partial^\mu\partial_\mu + m^2) A + B = 0,~~~~~~~
(\partial^\mu\partial_\mu + m^2) B = 0
\label{singleton}\eeq
these fields have the bulk $\mbox{AdS}$ action
\beq 
S = \int d^{D+1}x \sqrt{g}\left( g^{\mu\nu}\partial_\mu A \partial_\nu
B -m^2AB -\frac{1}{2}B^2 \right)
\label{action}\eeq
The fields  $A$ and $B$ couple to boundary fields $C$ and $D$ and
using eq. (\ref{adscft}) the two-point functions of $C$ and $D$ are
found to be (see \cite{kogan} for details of the calculation)
\bea
\langle C(x)C(y) \rangle &=& 0 \nonumber \\
\langle C(x)D(y) \rangle = \langle D(x)C(y) \rangle &=&
\frac{c}{|x-y|^2\Delta} \label{CD}\\
\langle D(x)D(y) \rangle &=& \frac{1}{|x-y|^2\Delta}(d-2c\ln|x-y|)
\nonumber \eea
with the dimension $\Delta$ given by $\Delta(\Delta-D)=m^2$, $c=
\Delta(2\Delta-D)$ and $d= 2\Delta - D$.
these are the usual two-point functions for logarithmic operators in
CFT \cite{gurarie,ckt}. These correlation functions occur if the
Hamiltonian (in two dimensions, the Virasoro generator $L_0$) is
non-diagonalizable, and has the Jordan form
\beq 
L_0 |C\rangle = h |C\rangle,~~~~~L_0 |D\rangle = h|D\rangle +
|C\rangle
\label{jordan} \eeq
and similarly for $\bar{L}_0$, where for singletons we have
$h=\bar{h}$ and so $\Delta = 2h$.
The theories with this type of operators are called Logarithmic CFTs
and their properties have been studied extensively \cite{ckt} since 
they were introduced in \cite{gurarie}. Applications of LCFT to
strings and D-brane scattering were developed in \cite{km,kmw}. A
recent paper relevant to $\mbox{AdS}_3$ is \cite{emw}

One way to see if fields of this type exist in
a theory is to look at the four-point functions of ordinary fields. If
there are no logarithmic operators, the operator product expansion
for primary fields has the form
\beq
O_i(x_1)O_j(x_2) \sim \sum_i
\frac{f^k_{ij}}{|x_{12}|^{\D_i+\D_j-\D_k}}O_k(x_1) + \cdots
\label{ope1}\eeq
and $\langle O_i(x_1)O_j(x_2) \rangle = |x_{12}|^{-2\D_i}\delta_{ij}$, 
which leads to an expansion for four-point functions of the form
\beq
\langle O_i(x_1)O_j(x_2)O_j(x_3)O_i(x_4) \rangle =
\sum_{kl}
\frac{f^k_{ij}f^l_{ij}}{|x_{12}|^{\D_i+\D_j}|x_{34}|^{\D_i+\D_j}} F(x)
\label{4point}\eeq
where $x=x_{12}x_{34}/x_{13}x_{24}$ and $F(x)$ has an expansion in
powers of $x$, without any logarithmic singularity. If there are
logarithmic operators however, the OPE has to be modified and we have
instead
\beq
O_i(x_1)O_j(x_2) \sim 
\frac{1}{|x_{12}|^{\D_i+\D_j-\D}}(D + C \ln |x_{12}|^2) + \cdots
\label{ope2}\eeq
which together with the two-point functions for $C$ and $D$ leads to
four point functions of the form (\ref{4point}), but with $F(x)
\stackrel{x \goto 0}{\goto} x^{2\D}\ln x$. Indeed, logarithmic
singularities have been found in four point functions calculated in
supergravity on $\mbox{AdS}_5$ \cite{hfmmr,sanjay}, 
and it is possible that these could be an
indication that there is an LCFT on the boundary of $\mbox{AdS}_5$. However,
these logarithms could also be accounted for as the perturbative
expansion of anomalous dimensions in $CFT_4$, with no need for
logarithmic operators \cite{bkrs}. The clearest evidence for the existence
of logarithmic operators in AdS/CFT comes from calculations of
grey-body factors in $\mbox{AdS}_3$. Since grey-body factors are related to
two-point functions in CFT, logarithms here are a clear indication
that we have logarithmic operators on the boundary.

The grey-body factor (or absorption cross section)
for a field in $\mbox{AdS}_3$ which couples to a field
$O(x)$ in the CFT on the boundary is related to the two
point function in the CFT by \cite{gubser,teo}
\beq
\sigma_{abs} = \frac{\pi}{\omega} \int d^2x 
\left[ {\cal G}(t-i\epsilon,x) - {\cal G}(t+i\epsilon,x) \right]
\label{gbf} \eeq
where ${\cal G}(t,x) = \langle O(x,t) O(0) \rangle$ is the thermal
Green's function in imaginary time. This can be determined from the
periodicity in imaginary time and the singularities of the Green's
function \cite{gubser}, which if $O$ is a primary field
with weights $h,\bar{h}$, are given by
\beq 
\langle O(t,x)O(0) \rangle \sim \frac{{\cal C}_O}{x_+^{2h}x_-^{2\bar h}}
\eeq
with $x_\pm = t \pm x$. ${\cal G}(t,x)$ has the form
\beq 
{\cal G}(t,x) = {\cal C} 
\left( \frac{\pi T_+R}{\sinh (\pi T_+x_+)}\right)^{2h}
\left( \frac{\pi T_-R}{\sinh (\pi T_-x_-)}\right)^{2\bar h}
\eeq
for a BTZ black hole with mass $M=r_+^2-r_-^2$, angular
momentum $J=2r_+r_-$, left and right temperatures $T_\pm=(r_+\pm
r_-)/2\pi$, and Hawking temperature given by $2/T_H=1/T_++1/T_-$  
\cite{btz}. The absorption cross section is then
\cite{gubser,teo}
\beq
\sigma_{abs}(h,\bar{h}) = \frac{\pi {\cal C}}{\omega} 
\frac{(2\pi T_+R)^{2h-1}(2\pi T_-R)^{2{\bar h}-1}}{\Gamma(2h) 
\Gamma(2\bar{h})}
\sinh\left(\frac{\omega}{2T_H}\right)
\left| \Gamma\left( h+i\frac{\omega}{4\pi T_+}\right)
 \Gamma\left( \bar{h}+i\frac{\omega}{4\pi T_-}\right) \right|^2
\label{gbf1}\eeq
This expression can be obtained either using the effective string
method for supergravity \cite{gubser} or using the Ads/CFT
correspondence \cite{teo,moz}.

A large number of classical calculations of absorption cross sections
have given results which are consistent with (\ref{gbf1}) (or a
similar expression for fermions) \cite{gubser}, including calculations
for several fields for the BTZ black hole \cite{klm,teo}. However, in
\cite{klm} the
cross section for gauge bosons with spin $2$, which couple to fields
with $h,\bar{h} = (2,0)$ or $(0,2)$ on the boundary was found to have
logarithmic corrections which cannot be accounted for by
eq. (\ref{gbf1}). In \cite{ml}, the grey-body factor for singletons
was calculated, and while this does have a logarithmic correction to
the cross section (\ref{gbf1}), it was found that it still does not
give the correct  cross section for the gauge bosons. The question we
would like to address in this letter is, are there other kinds of
logarithmic operators in $\mbox{AdS}_3/\mbox{CFT}_2$, and can 
they correctly account
for the greybody factor for the gauge bosons?

The greybody factor for the gauge bosons with spin $s=2$
in $\mbox{AdS}_3$, in the low
temperature limit $\omega \gg T_\pm$, was found to be \cite{klm}
\beq 
\sigma_{abs}^{gb} = \pi^2 \omega R^2 \left[ 1+\omega Rs 
\ln (2\omega Rs) \right]
\label{gbfgb}\eeq
In the low temperature limit, eq. (\ref{gbf1}) becomes, up to a
normalization whch is proportional to $\cal C$
($\Delta=h+\bar{h}$),
\beq 
\sigma_{abs}(h,\bar{h}) \sim 
\omega^{2\Delta -3}
\eeq
So that the second term in eq. (\ref{gbfgb}) is an indication that the
gauge bosons cannot just couple to ordinary primary fields on the boundary. 
The greybody factor for a singleton can also be calculated from
eq. (\ref{gbf}), using the relation $\langle D(t,x)D(0) \rangle
=\frac{\partial}{\partial\Delta} \langle C(t,x)D(0) \rangle$, since 
$\langle C(t,x)D(0) \rangle$ is the same as the two point function for
an ordinary primary field. The greybody factor for a singleton is
therefore given by $\sigma_{abs}^S =
\partial\sigma_{abs}(h,\bar{h})/\partial\Delta$ \cite{ml}, and so
\bea
\sigma_{abs}^S &=& \frac{\pi {\cal C}}{\omega} 
\frac{(2\pi T_+R)^{2h-1}(2\pi T_-R)^{2{\bar h}-1}}{\Gamma(2h) 
\Gamma(2\bar{h})}
\sinh\left(\frac{\omega}{2T_H}\right) 
\left| \Gamma\left( h+i\frac{\omega}{4\pi T_+}\right)
 \Gamma\left( \bar{h}+i\frac{\omega}{4\pi T_-}\right) \right|^2
\nonumber \\ &\times&
\left[\frac{1}{\cal C}\frac{\partial  \cal C} {\partial\Delta}
+\ln(2\pi T_+ R)+\ln(2\pi T_- R) -\psi(h)-\psi(2\bar{h})\right.\\
&&\left.+\frac{1}{2}\left\{ 
\psi\left(h+i\frac{\omega}{4\pi T_+}\right)+
\psi\left(h-i\frac{\omega}{4\pi T_+}\right)+
\psi\left(\bar{h}+i\frac{\omega}{4\pi T_+}\right)+
\psi\left(\bar{h}-i\frac{\omega}{4\pi T_+}\right)
\right\}\right]
\nonumber\eea
Which in the low temperature limit reduces to
\beq
\sigma_{abs}^S \sim \omega^{2\Delta-3}\left( 2\ln(\omega R) +c'
\right)
\label{gbfsing}\eeq
In an LCFT we always have the freedom to shift $D$ by $D \goto D +
\lambda C$, which leaves eq. (\ref{jordan}) invariant, and this can be
used to choose any value for the constant $c'$. However, comparing
eqs. (\ref{gbfsing}) and (\ref{gbfgb}), we can see that the
logarithmic term in (\ref{gbfgb}) is multiplied by an extra factor of
$\omega R$ and is thus of a sub-leading order compared to
(\ref{gbfsing}). The gauge boson cannot therefore be represented by a
singleton in $\mbox{AdS}_3$ \cite{ml}. However, we cannot immediately
conclude, as was said in \cite{ml}, that the gauge boson has nothing
to do with the AdS/LCFT correspondence, because there is potentially a
much richer spectrum of logarithmic operators in a two dimensional
LCFT than has been considered so far. The logarithmic operators we
have looked at so far arise when the dimensions of
two of the primary fields $O_k$ in
the OPE (\ref{ope1}) become degenerate, which leads to logarithms in
the four-point functions and the OPE has to be modified to include the
logarithmic pair $C$ and $D$, as in eq. (\ref{ope2}). In fact,
logarithms will also arise in the four point function if two of the
primary fields have dimensions which are not equal, but differ by an
integer, so that it is a descendant of one primary field which becomes
degenerate with the other primary field. This is because the function
$F(x)$ in the four-point function (\ref{4point}) usually satisfies a
Fuchsian differential equation, such as a hypergeometric equation, and
when there are no degenerate dimensions the solutions have the form
\beq 
F(x) \sim x^{\Delta_i}\sum_{n=0}^\infty a_nx^n
\eeq
but when two of the dimensions differ by an integer, say
$\Delta_2=\Delta_1+N$, the second solution instead has the form
\beq
F(x) \sim x^{\Delta_i}\sum_{n=0}^\infty (a_nx^n+b_nx^n\log x)
\eeq
in this case we again have a logarithmic pair with the higher of the
two dimensions which as before make the Hamiltonian
non-diagonalizable, as in eq. (\ref{jordan}). In addition the $C$
field satisfies in both cases the usual condition for a primary field
\beq
L_n|C\rangle =0,~~~~~n\geq 1
\label{primary}\eeq
However, in the earlier situation where two primary fields became
degenerate, $D$ also satisfied this condition, while in the case where
we have two fields whose dimension differs by an integer $N$ we have
instead 
\beq
(L_1)^N|D\rangle = \beta|C'\rangle,~~~~~L_n|D\rangle =0,~~~~~n\geq 2
\eeq
where $C'$ is another primary field, with conformal weights
$(h-N,\bar{h})$, and $\beta$ is some constant. 
$C$ is now not really a primary field, but rather a
descendant of $C'$:  $|C\rangle = \sigma_{-N}|C'\rangle$, where
$\sigma_{-N}$ is some combination of Virasoro generators and, in general,
the other generators of the chiral algebra of the CFT, of dimension
$N$. Eq. (\ref{primary}) then implies that $C$ must be a null vector
of the CFT,
that is $[L_n,\sigma_{-N}]=0$ for $n\geq 1$ \cite{rohsiepe}
(which is why the
two-point function $\langle CC \rangle =0$). 
This type of logarithmic
operator therefore cannot exist with any dimension, but only with those
dimensions for which there are null vectors of the algebra. Because of
this, we can only have these generalized logarithmic operators in
$2$-dimensional CFT, and we do not expect them in $\mbox{AdS}_{D+1}$ for $D>2$.

The logarithmic pair $C$ and $D$ still have the same correlation
functions (\ref{CD}), and $C'$ is just an ordinary primary field with
the usual two point function
\beq 
\langle C'(x_+,x_-)C'(-0)\rangle \sim
\frac{1}{x_+^{2(h-N)}x_-^{2\bar{h}}}
\eeq
so it seems that these new fields cannot give us anything new when we
compute greybody factors. However, it is easy to see that we can
reproduce the greybody factor for the gauge bosons (\ref{gbfgb}) if we
assume that they correspond not to one of the fields $C$, $D$ or $C'$ 
in the
LCFT, but to a linear combination of all three. This might happen, for
example, if the bosons can be thought of as arising from the fusion of
two primary fields, since $C$, $D$ and $C'$ must always appear
together in any OPE. Then if $C'$ has dimension$\Delta=2$, as is
expected for the gauge bosons \cite{ml}, and $C',C,D$ form a
representation of the type discussed above with $N=1$, the greybody
factor will have exactly the right form, with the logarithmic term
being of sub-leading order. Of course, this would imply that the
representation which includes the primary field $C'$ must have a null
vector at level $1$. This would be true if $C'$ has
$(h,\bar{h})=(0,2)$ (or $(2,0)$), as then $L_{-1}|C'\rangle$ (or 
$\bar{L}_{-1}|C'\rangle$) is a null vector. If $C'$ has
$(h,\bar{h})=(1,1)$, there could still be a null; vector if, for
example, the CFT on the boundary has a conserved current for which
$J_{-1}|C'\rangle=0$.

Now that we know there could be fields in an LCFT on the boundary that
give the correct greybody factor for the gauge bosons, the next
question we address is, what sort of fields in the bulk can couple to
these fields on the boundary? To answer this question, we start be
reviewing how the conformal weights of fields on the boundary
determine the mass and spin of fields in the bulk when there are no
logarithmic operators. We write the metric for $\mbox{AdS}_3$ in the form
\beq
ds^2 = l^2\left( -\cosh^2\rho d\tau^2 + \sinh^2\rho d\phi^2 + d\rho^2
\right)
\label{metric}\eeq
In these coordinates the Virasoro generators $L_0,L_{\pm 1}$, with
commutators $[L_0,L_{\pm 1}]=\mp L_{\pm 1}$ and $[L_1,L_{-1}]=2L_0$,
 for spin
$s$ fields are
($u=\tau+\phi,~v=\tau-\phi$) \cite{ms,deboer}:
\bea
L_0 &=& i\partial_u \nonumber \\
L_{-1} &=& ie^{-iu}\left(\coth 2\rho \partial_u -\frac{1}{\sinh
2\rho}\partial_v +\frac{i}{2}\partial_\rho - \frac{i}{2}s\coth\rho
\right) \nonumber \\
L_{1} &=& ie^{iu}\left(\coth 2\rho \partial_u -\frac{1}{\sinh
2\rho}\partial_v -\frac{i}{2}\partial_\rho + \frac{i}{2}s\coth\rho
\right)
\label{virs}\eea
and similarly for $\bar{L}_0,\bar{L}_{\pm 1}$ with $u \leftrightarrow
v$ and $s \goto -s$. For a primary fields $\Phi$, the conditions
$L_0 \Phi=h\Phi$ and
$\bar{L}_0 \Phi=\bar{h}\Phi$, and $L_1\Phi=\bar{L}_1\Phi=0$ can then
be solved to give $s=h-\bar{h}$ and
\beq
\Phi \sim \frac{e^{-i(hu+\bar{h}v)}}{(\cosh\rho)^{h+\bar{h}}}
\eeq
The second Casimir of  $sl(2,{\bf R})$ is
\beq
L^2 = \frac{1}{2}(L_1L_{-1} + L_{-1}L_1) -L_0^2
\eeq
and similarly for $\bar{L}^2$, so that, using eqs. (\ref{metric}) and
(\ref{virs}),  the sum of the two Casimirs is
\beq
L^2 + \bar{L}^2 = -l^2\partial^\mu\partial_\mu +s^2\coth^2\rho
\label{casimir}\eeq
For a primary field, $(L^2+\bar{L}^2)\Phi =
(2h(h-1)+2\bar{h}(\bar{h}-1))\Phi$, so eq. (\ref{casimir}) can be
written as
\beq
\left( -\partial^\mu\partial_\mu + \frac{s^2}{l^2\sinh^2\rho}
\right)\Phi 
= m^2\Phi
\eeq
which is the equation of motion for a field with spin $s$ and mass $m$
in $\mbox{AdS}_3$, with the mass
\beq
l^2m^2 = 2h(h-1) + 2\bar{H}(\bar{h}-1) -s^2 = \Delta(\Delta-2)
\label{mass}\eeq
So we can see that the conformal weights $h$ and $\bar{h}$ on the
boundary completely determine the mass and spin of the fields in
$\mbox{AdS}_3$ (and vice versa). We can repeat this analysis for a
logarithmic pair $C$ and $D$ on the boundary. $C$ satisfies the same
conditions as $\Phi$ above, so we find $C\sim
e^{-i(hu+\bar{h}v)}/(\cosh\rho)^{h+\bar{h}}$. The conditions
$L_0D=hD+C$ and $\bar{L}_0D=\bar{h}D+C$ then imply that
\beq
D=[u+v+f(\rho)]C
\label{frho}\eeq
where the function $f(\rho)$ will depend on what type of logarithmic
operator we have. In the simplest case, which we expect to give us
singletons, we have $L_1D= \bar{L_1}D=0$, which has the solution
\beq
D=[u+v-2i\ln(\cosh\rho)+\delta]C
\eeq
where $\delta$ is an arbitrary constant, which we can set to any value
using the freedom to shift $D$ by an amount proportional to $C$.
Evaluating the second Casimirs gives the equations of motion for
the fields in $\mbox{AdS}_3$ which will couple to $C$ and $D$:
\bea
\left( -\partial^\mu\partial_\mu + \frac{s^2}{l^2\sinh^2\rho}
\right)C    
&=& m^2C \nonumber \\
\left( -\partial^\mu\partial_\mu + \frac{s^2}{l^2\sinh^2\rho}
\right)D    
&=& m^2D + 4(\Delta-1)C
\eea 
with $m^2$ again given by eq. (\ref{mass}). When $s=0$, these are just
the expected equations of motion for singleton dipole-pair
(\ref{singleton}) (apart from a different normalization of $C$), with
the expected relation between the singleton mass $m$ and the dimension
of the logarithmic operator $\Delta$. Thus we can see that the mass and
spin in $\mbox{AdS}_3$ are still completely determined by the data of the LCFT
on the boundary when we have singletons and logarithmic
operators. This also give us another way of seeing, as was found in
\cite{kogan} that there can be no logarithmic operators with
$\Delta=1$, corresponding to $m^2=-1$.  

Now we can use this map between $\mbox{AdS}_3$ and $\mbox{CFT}_2$ to see
what kind of operators will couple to the other kinds of logarithmic
operators. In this case we have three fields $C,D$ and $C'$, but since
$C$ and $C'$ are both primary fields they will both have the same
form as before, but with weights $(h,\bar{h})$ for $C$ and
$(h-N,\bar{h})$ for $C'$. 
$D$ is then given by  eq. (\ref{frho}) with $f(\rho)$ a solution of
the $(N+1)$'th order differential equation 
$(L_1)^{N+1}D=0$.
The second casimirs then give the equations of motion in $\mbox{AdS}_3$ as
\bea
\left( -\partial^\mu\partial_\mu + \frac{s^2}{l^2\sinh^2\rho}
\right)C    
&=& m^2C \nonumber \\
\left( -\partial^\mu\partial_\mu + \frac{s^2}{l^2\sinh^2\rho}
\right)D    
&=& m^2D + 4(\Delta-1)C +\Psi
\eea
Which is the same as the equations of motion for the singleton except
 that we have the new  field is $\Psi=L_{-1}L_1D$. 
$\Psi$ will therefore be a
 descendant of $C'$, or in $\mbox{AdS}_3$ it will correspond to some  derivative
 of the field which couples to the primary field $C'$, 
and the action for the singleton (\ref{action})
 should be modified by adding a term which couples the singleton to
 the new field. We will therefore have an interacting theory instead
 of a free singleton, with an action of the form
\beq 
S = \int d3x \sqrt{g}\left( g^{\mu\nu}\partial_\mu A \partial_\nu
B -m^2AB -\frac{1}{2}B^2 +\lambda A\Psi\right)
\eeq
where $\Psi$ is a derivative of a field with spin $N$, for a spinless
singleton. In addition, it is important that $C$ is also a descendant
in this case, and so the field $A$ in the above action is also a
derivative of the field which couples to $C'$ and is not a fundamental
field itself. This is especially significant for the case when $N=1$,
since then $C'$ has no descendant at level $1$ except $C$ itself, and
so the action in $\mbox{AdS}_3$ in this case is the same as for the ordinary
singleton, {\it except} that $B$ is now a derivative of a field $B'$ with
spin $1$. Of course, we also need to add to
the action the kinetic and mass terms for the field $B'$ to
treat this field properly.

Although we have seen that new kinds of logarithmic operators can
exist in $\mbox{AdS}_3/\mbox{LCFT}_2$, 
they cannot exist for just any values of
$m^2$ and $s$ - we have to have a null vector in the CFT on the
boundary. This means that to determine if such fields really exist we
need to know more about the structure of the CFT, or to calculate
four-point functions, from which the OPE could be deduced. However, it
seems to be clear that at least one example of this type of operator
is needed to give the correct greybody factor for the spin-$2$ gauge
bosons. It is an interesting question why the same interactions cannot
be introduced for singletons which do not have special values for the
mass, which would lead to a contradiction in the CFT, but is not
obviously forbidden from the three-dimensional point of view. Possibly
related is the question of why these type of fields can exist in
$\mbox{AdS}_3$ but not in $\mbox{AdS}_{D+1}$ for $D>2$ -- since the full Virasoro
algebra applies only to CFT in 2 dimensions, there are no null vectors
in $D>2$ and so these type of logarithmic do not exist, although there
can be singletons and the ordinary logarithmic pair of $C$ and $D$ in
any dimension.

\section{Acknowledgments}

This work was supported by Enterprise Ireland grant no. SC/98/739.

\end{document}